\documentclass[twocolumn,showpacs,preprintnumbers,amsmath,amssymb,prl,superscriptaddress]{revtex4}

\usepackage{graphicx} 
\usepackage{dcolumn}  
\usepackage{colordvi}

\graphicspath{{ps}}

\newcommand{\x}{$\times$}

\begin{document}

\preprint{\vbox{ \hbox{   }
    \hbox{Belle Preprint \# 2011-15}
    \hbox{KEK Preprint \# 2011-19}
}}

\title { \quad\\[0.5cm] Search for $CP$ Violation in $D$ Meson Decays to $\phi\pi^+$}

\date{\today}

\begin{abstract}
  We search for $CP$ violation in Cabibbo-suppressed charged $D$ meson decays 
  by measuring the  
  difference between the $CP$ violating asymmetries for the Cabibbo-suppressed decays
  $D^+ \to K^+K^- \pi^+$ and the Cabibbo-favored decays $D_s^+ \to K^+K^- \pi^+$ 
  in the $K^+K^-$ mass region of the $\phi$ resonance. Using 955~fb$^{-1}$ of
  data collected with the Belle detector 
  we obtain $A_{CP}^{D^+ \to \phi \pi^+} = (+0.51 \pm 0.28 \pm 0.05)\%$. 
  The measurement improves the sensitivity of previous searches by more
  than a factor of five. We find no evidence for direct $CP$ violation.
\end{abstract}

\affiliation{University of Bonn, Bonn}
\affiliation{Budker Institute of Nuclear Physics SB RAS and Novosibirsk State University, Novosibirsk 630090}
\affiliation{Faculty of Mathematics and Physics, Charles University, Prague}
\affiliation{University of Cincinnati, Cincinnati, Ohio 45221}
\affiliation{Justus-Liebig-Universit\"at Gie\ss{}en, Gie\ss{}en}
\affiliation{Gifu University, Gifu}
\affiliation{Hanyang University, Seoul}
\affiliation{University of Hawaii, Honolulu, Hawaii 96822}
\affiliation{High Energy Accelerator Research Organization (KEK), Tsukuba}
\affiliation{Indian Institute of Technology Guwahati, Guwahati}
\affiliation{Indian Institute of Technology Madras, Madras}
\affiliation{Indiana University, Bloomington, Indiana 47408}
\affiliation{Institute of High Energy Physics, Chinese Academy of Sciences, Beijing}
\affiliation{Institute of High Energy Physics, Vienna}
\affiliation{Institute of High Energy Physics, Protvino}
\affiliation{Institute for Theoretical and Experimental Physics, Moscow}
\affiliation{J. Stefan Institute, Ljubljana}
\affiliation{Kanagawa University, Yokohama}
\affiliation{Institut f\"ur Experimentelle Kernphysik, Karlsruher Institut f\"ur Technologie, Karlsruhe}
\affiliation{Korea Institute of Science and Technology Information, Daejeon}
\affiliation{Korea University, Seoul}
\affiliation{Kyungpook National University, Taegu}
\affiliation{\'Ecole Polytechnique F\'ed\'erale de Lausanne (EPFL), Lausanne}
\affiliation{Faculty of Mathematics and Physics, University of Ljubljana, Ljubljana}
\affiliation{University of Maribor, Maribor}
\affiliation{Max-Planck-Institut f\"ur Physik, M\"unchen}
\affiliation{University of Melbourne, School of Physics, Victoria 3010}
\affiliation{Nagoya University, Nagoya}
\affiliation{Nara Women's University, Nara}
\affiliation{National Central University, Chung-li}
\affiliation{National United University, Miao Li}
\affiliation{Department of Physics, National Taiwan University, Taipei}
\affiliation{H. Niewodniczanski Institute of Nuclear Physics, Krakow}
\affiliation{Nippon Dental University, Niigata}
\affiliation{Niigata University, Niigata}
\affiliation{University of Nova Gorica, Nova Gorica}
\affiliation{Osaka City University, Osaka}
\affiliation{Pacific Northwest National Laboratory, Richland, Washington 99352}
\affiliation{Panjab University, Chandigarh}
\affiliation{Research Center for Nuclear Physics, Osaka}
\affiliation{University of Science and Technology of China, Hefei}
\affiliation{Seoul National University, Seoul}
\affiliation{Sungkyunkwan University, Suwon}
\affiliation{School of Physics, University of Sydney, NSW 2006}
\affiliation{Tata Institute of Fundamental Research, Mumbai}
\affiliation{Excellence Cluster Universe, Technische Universit\"at M\"unchen, Garching}
\affiliation{Tohoku Gakuin University, Tagajo}
\affiliation{Tohoku University, Sendai}
\affiliation{Department of Physics, University of Tokyo, Tokyo}
\affiliation{Tokyo Institute of Technology, Tokyo}
\affiliation{Tokyo Metropolitan University, Tokyo}
\affiliation{Tokyo University of Agriculture and Technology, Tokyo}
\affiliation{CNP, Virginia Polytechnic Institute and State University, Blacksburg, Virginia 24061}
\affiliation{Yonsei University, Seoul}
  \author{M.~Stari\v{c}}\affiliation{J. Stefan Institute, Ljubljana} 
  \author{H.~Aihara}\affiliation{Department of Physics, University of Tokyo, Tokyo} 
  \author{K.~Arinstein}\affiliation{Budker Institute of Nuclear Physics SB RAS and Novosibirsk State University, Novosibirsk 630090} 
  \author{D.~M.~Asner}\affiliation{Pacific Northwest National Laboratory, Richland, Washington 99352} 
  \author{T.~Aushev}\affiliation{Institute for Theoretical and Experimental Physics, Moscow} 
  \author{A.~M.~Bakich}\affiliation{School of Physics, University of Sydney, NSW 2006} 
  \author{A.~Bay}\affiliation{\'Ecole Polytechnique F\'ed\'erale de Lausanne (EPFL), Lausanne} 
  \author{V.~Bhardwaj}\affiliation{Panjab University, Chandigarh} 
  \author{B.~Bhuyan}\affiliation{Indian Institute of Technology Guwahati, Guwahati} 
  \author{A.~Bozek}\affiliation{H. Niewodniczanski Institute of Nuclear Physics, Krakow} 
% ...
\collaboration{The Belle Collaboration}

\maketitle

Studying $CP$ asymmetries in $D$ meson decays provides a promising
opportunity to search for new physics (NP) beyond the Standard Model
(SM)~\cite{Bianco}. Here we study $CP$ asymmetries in charged
$D^+ \to \phi\pi^+$ and $D^+_s \to \phi\pi^+$ decays~\cite{ch_conj}. 
The observable of interest is
\begin{equation}
  \label{a_f}
  A_{CP}^{D_{(s)}^+ \to \phi \pi^+} =
  \frac{\Gamma(D_{(s)}^+ \to \phi \pi^+) - \Gamma(D_{(s)}^- \to \phi \pi^-)}
       {\Gamma(D_{(s)}^+ \to \phi \pi^+) + \Gamma(D_{(s)}^- \to \phi \pi^-)}~,
\end{equation}
where $\Gamma$ is the partial decay width. 
This time-integrated asymmetry arises from $CP$ violation (CPV)
in decay amplitudes. Within the SM, CPV in $D$ decay amplitudes 
is predicted to be very small.
The largest effect occurs for singly Cabibbo-suppressed (SCS)
decays such as $D^+ \to \phi\pi^+$, which are governed by the Cabibbo-Kobayashi-Maskawa (CKM)  
matrix elements $V_{cs}V^\ast_{us}$. However, even for these decays $A_{CP}$ is
predicted to be only ${\cal{O}}(0.1\%)$ or less ~\cite{Grossman}. In contrast, several NP 
models predict $A_{CP}$ to be as large as~1\%. 
Experimentally, to cancel detector-induced asymmetries and other systematic
effects, we measure the difference 
\begin{equation}
  \label{Arec.def}
  \Delta A_{\rm rec} =\frac{N(D^+) - N(D^-)}{N(D^+) + N(D^-)}-
  \frac{N(D_s^+) - N(D_s^-)}{N(D_s^+) + N(D_s^-)}~,
\end{equation}
where the second term corresponds to the Cabibbo-favored (CF)
decay $D^+_s \to \phi\pi^+$. This CF decay is governed by the
CKM matrix elements $V^{}_{cs}V^\ast_{ud}$ and is expected to have 
negligible $A_{CP}$~\cite{bergmann}; thus, a measurement
of $\Delta A_{\rm rec}$ probes $A_{CP}^{D^+ \to \phi \pi^+}$. Measuring a relatively
large value would be interpreted as evidence for NP.
Previously, CPV in $D$ meson SCS decays has been searched for 
in several final states~\cite{CPV_dir}. No significant asymmetries
were found, with the best sensitivities ranging from ${\cal{O}}(0.2\%)$ to 
${\cal{O}}(2\%)$ depending on the decay mode~\cite{BRko,CPV_old,StaricPLB}.  

The measurement is based on 955~fb$^{-1}$ of data recorded
with the Belle detector~\cite{Belle} at the KEKB asymmetric-energy
$e^+e^-$~collider~\cite{KEKB}, which primarily operated at the center-of-mass (CM) energy of
the $\Upsilon(4S)$ resonance and 60~MeV below. A fraction of the data was recorded 
at the $\Upsilon(1S)$, $\Upsilon(2S)$, $\Upsilon(3S)$, and $\Upsilon(5S)$ resonances; 
these data are included in the measurement. 
The Belle detector is described in detail elsewhere~\cite{Belle,SVD2}:
in particular, it includes a silicon vertex detector (SVD), 
a central drift chamber,
an array of aerogel Cherenkov counters,
time-of-flight scintillation counters,
an electromagnetic calorimeter and a muon detector.

We reconstruct the decays $D_{(s)}^+ \to \phi \pi^+$ in the $\phi \to K^+K^-$ decay mode. 
Each final state charged particle is required to have at least two associated SVD hits 
in each of the two measured coordinates. To select pion and kaon candidates, we impose
standard particle identification criteria~\cite{PID}. The identification efficiencies
and the misidentification probabilities are about 90\% and 5\%, respectively. In addition, 
we require loose proton veto criteria for kaon candidates and loose lepton veto 
criteria for pion candidates, since we found that a considerable fraction of background (23\%) 
involves misidentified protons and leptons. $D$ meson daughter particles are refitted to
a common vertex and the $D$ meson candidate is constrained to originate from the $e^+e^-$
interaction region. Confidence levels exceeding 10$^{-3}$ are required for both fits. 
In order to reject $D$ mesons produced in $B$ meson decays, the $D$ meson momentum in 
the $e^+e^-$ CM system must satisfy $p^*_D > 2.5~{\rm GeV}/c$, 
for the data taken below $\Upsilon(5S)$, and $p^*_D > 3.1~{\rm GeV}/c$ for the 
$\Upsilon(5S)$ data.

We accept candidates in the invariant mass regions of $D$ and $\phi$ mesons,
$1.80~{\rm GeV}/c^2 < M_{KK\pi} < 2.05~{\rm GeV}/c^2$ and $M_{KK} < 1.07~{\rm GeV}/c^2$. 
For the small fraction of events with multiple candidates (4.6\%), we select a single best 
candidate: the one with the smallest $\chi^2$ of the production and decay vertex fits.

We study background using a generic Monte Carlo (MC) simulation 
based on EVTGEN~\cite{evtgen} and GEANT3~\cite{geant3}.
We find that the main component (97\%) is the combinatorial background 
whose shape in $M_{KK\pi}$ can be fitted well with an exponential function. 
Other background components are mainly due to decays of charm particles and 
have a complicated structure in $M_{KK\pi}$.
However, their fractions are sufficiently small that the structure is obscured by the
statistical fluctuations of the main background component. The combinatorial background can be 
further divided into random combinations of a correctly reconstructed $\phi$ meson 
and a $\pi^+$ (42\%), and the rest (58\%).

To improve the purity of the $D^+$ and $D_s^+$ data sample we require: 
$|M_{KK}-m_\phi| < 16~{\rm MeV}/c^2$, where $m_\phi$ is the nominal mass of $\phi$,
$p_\pi > 0.38~{\rm GeV}/c$, where $p_\pi$ is the laboratory momentum of $\pi^+$, and
$|\cos \theta_{\rm hel}| > 0.28$, where $\theta_{\rm hel}$ is the angle
between $K^-$ and $D_{(s)}^+$ momenta in the rest frame of $\phi$.
These selection criteria are obtained by minimizing the expected statistical error
on $\Delta A_{\rm rec}$ using signal and background samples from the generic MC simulation.
The simulation has been tuned prior to running the optimization procedure to reproduce 
the mass resolutions of the $D$ signals in data 
and the signal-to-background ratios of the data.

\begin{figure}[t]
  \includegraphics[width=8.5cm]{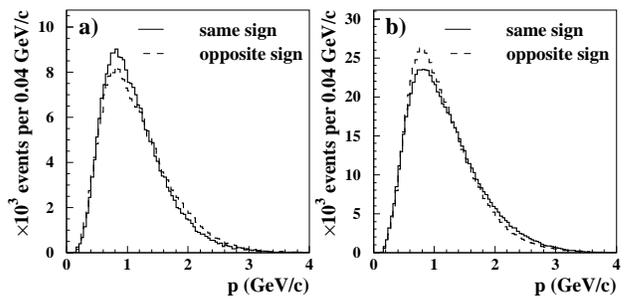}
  \caption{Background subtracted momentum distributions of kaons
    of the same and opposite charges relative to that of the $D$ meson:
    (a) for $D^\pm$ decays and (b) for $D_s^\pm$ decays. 
    Background is taken from sidebands in $M_{KK\pi}$.}
  \label{fig1}
\end{figure}

The measured asymmetry $A_{\rm rec}$ 
can be written as the sum of several contributions that are assumed to be small:
\begin{equation}
  \label{Arec.eq}
  A_{\rm rec} = A_{CP} + A_{FB}(\cos \theta^*) + A_\epsilon^{KK} +
  A_\epsilon^\pi(p_\pi, \cos \theta_\pi).
\end{equation}
In addition to the intrinsic asymmetry, $A_{CP}$, there is a forward-backward asymmetry 
($A_{FB}$) in the production of $D$ mesons in $e^+e^- \to c\overline{c}$ arising from 
$\gamma-Z^0$ interference and higher-order QED effects. This term is an odd function 
of the cosine of the $D$ meson production polar angle $\theta^*$ in the CM system and 
could differ between $D^+$ and $D_s^+$ due to fragmentation effects. 
Furthermore, there are contributions due to asymmetry in the reconstruction efficiencies 
of oppositely charged kaons ($A_\epsilon^{KK}$) and pions ($A_\epsilon^\pi$). 
The term $A_\epsilon^{KK} \equiv 0$ for $\phi \to K^+K^-$ decays. However, the
interference with other intermediate states in the decay $D_{(s)}^+ \to K^+K^-\pi^+$
introduces a small difference in momentum distributions of the
same-sign and the opposite-sign kaons (where the sign is relative to the $D_{(s)}$ 
meson charge), as shown in Fig.~\ref{fig1}. 
The difference in momentum distributions in combination with the kaon detection asymmetry 
$A_\epsilon^K$ leads to a non-zero $A_\epsilon^{KK}$, as explained below.

We define the intrinsic laboratory phase space distribution
of the kaon pair by $P(x_1,x_2)$,
where $x_1 \equiv (p_1,\cos \theta_1)$ and $x_2 \equiv (p_2,\cos \theta_2)$
label the phase space variables of same-sign and opposite-sign kaons, respectively.
The measured same-sign and opposite-sign single kaon distributions are obtained
from the intrinsic $P(x_1,x_2)$ by:
\begin{equation}
  \label{P1.eq}
  P_{1(2)}(x_{1(2)}) = \frac{\epsilon(x_{1(2)}) \int P(x_1,x_2) \epsilon(x_{2(1)}) dx_{2(1)}}
  {\int \int dx_1 dx_2 P(x_1,x_2) \epsilon(x_1) \epsilon(x_2)}~, 
\end{equation}
where $\epsilon(x)$ is the phase-space dependent detection efficiency.
The numbers of detected positively and negatively charged $D$ mesons are
\begin{eqnarray}
  N^\pm = \int \int dx_1 dx_2 P(x_1,x_2) \epsilon_{K^\pm}(x_1) \epsilon_{K^\mp}(x_2),
\end{eqnarray}
where $\epsilon_{K^\pm}(x) = \epsilon(x)(1\pm A_\epsilon^K(x))$
are the efficiencies of the $K^\pm$ as functions of kaon phase space $x$.
From this and neglecting terms quadratic in $A_\epsilon^K$ one obtains:
\begin{equation}
  \label{A_KK.eq}
  A_\epsilon^{KK} = \int (P_1(x) - P_2(x)) A_\epsilon^K(x) dx,
\end{equation}
where $P_1(x)$ and $P_2(x)$ are normalized distributions of the detected
same-sign and opposite-sign kaons, respectively, given by Eq.~(\ref{P1.eq}) 
and the integration runs over the kaon phase space $x \equiv (p, \cos \theta)$.

The last term in Eq.~(\ref{Arec.eq}) is a function 
of pion momentum and polar angle in the laboratory frame. 
In the difference of measured $D^+$ and $D_s^+$ asymmetries,
provided the measurement is done in bins of the three-dimensional (3D) phase space
$(\cos \theta^*, p_\pi, \cos \theta_\pi)$, the last term in Eq.~(\ref{Arec.eq}) cancels,
\begin{equation}
  \label{D_Areco}
  \Delta A_{\rm rec} = A_{CP}^{D^+ \to \phi \pi^+} 
  + \Delta A_{FB}(\cos \theta^*) + \Delta A_\epsilon^{KK}.
\end{equation}

In the above equation we assume that the intrinsic $A_{CP}^{D_s^+ \to  \phi \pi^+}$ 
is negligible, as discussed in the introduction. 
We use 10\x 10\x 10 equal size bins of the 3D phase space with $p_\pi < 5~{\rm GeV}/c$.
The yields of $D^+$, $D^-$, $ D_s^+$ and $D_s^-$ decays are
determined from a binned likelihood fit to the $M_{KK\pi}$ distributions in each sufficiently
populated 3D bin. We require at least 100 entries
in the histogram in order to perform the fit. To parameterize the non-Gaussian signal shape 
with as few parameters as possible, we use the distribution of pulls determined with 
MC simulation; the pulls are calculated as $(M_{KK\pi}-\overline{m})/\sigma_m$, 
where $\overline{m}$ and $\sigma_m$ are the 3D bin dependent mean and standard deviation 
of the $D^+$ or $D_s^+$ invariant mass distributions.
The pull distribution is fitted with a sum of four Gaussians
to obtain their fractions $f_i^{\rm pull}$, mean positions $x_i^{\rm pull}$ and
the widths $\sigma_i^{\rm pull}$.
The signal shape for the decays with no final state radiation (FSR) is parameterized with:
\begin{equation}
  \label{Sg.eq}
  S_{4g}(x)=\sum_{i=1}^4 \frac{f_i^{\rm pull}}{\sqrt{2\pi}s_i} e^{-\frac{(x-x_i^0)^2}{2s_i^2}},
\end{equation}
where $s_i = \sigma_i^{\rm pull} \sigma$ and 
$x_i^0 = x_i^{\rm pull} \sigma_i^{\rm pull}+ x_0$.
The normalized shape given by Eq.~(\ref{Sg.eq}) has two free varying parameters:
the position $x_0$ and the width $\sigma$.
 
The pull distribution is found to be $\cos \theta^*$ dependent;
it becomes asymmetric at the edges of $\cos \theta^*$ space. Thus, to improve the
mass fits, we use $\cos \theta^*$ dependent pull parameters  $f_i^{\rm pull}$,
$x_i^{\rm pull}$ and $\sigma_i^{\rm pull}$ in Eq.~(\ref{Sg.eq}), which are obtained from
fits to simulated distributions in ten bins of $\cos \theta^*$.

The shape of the FSR tail $S_{\rm FSR}(x)$ and its fraction $p_{\rm FSR}$~\cite{p_FSR} are taken from
MC simulation using PHOTOS~\cite{photos} to simulate FSR. The normalized signal shape is then 
$S(x) = (1-p_{\rm FSR}) S_{4g}(x) + p_{\rm FSR} S_{\rm FSR}(x)$. 
The background is parameterized with an exponential function of a quadratic polynomial,
$B(x)=e^{a+bx+cx^2}$, with free parameters $a$, $b$, and $c$. 

The parameterization of the $M_{KK\pi}$ distribution includes two
signal peaks and the background, and has eight free varying parameters:
two yields ($N_{D^+}$, $N_{D_s}$), two peak positions ($m_{D^+}$, $m_{D_s}$),
the width of $D^+$ peak ($\sigma_{D^+}$) and three background parameters
($a$, $b$, $c$). The ratio of the $D_s^\pm$ and $D^\pm$ peak widths,
$f=\sigma_{D_s}/\sigma_{D^+}$, is fixed from MC simulation in order to ensure stable fitting.

\begin{figure}[t]
  \includegraphics[width=8.5cm]{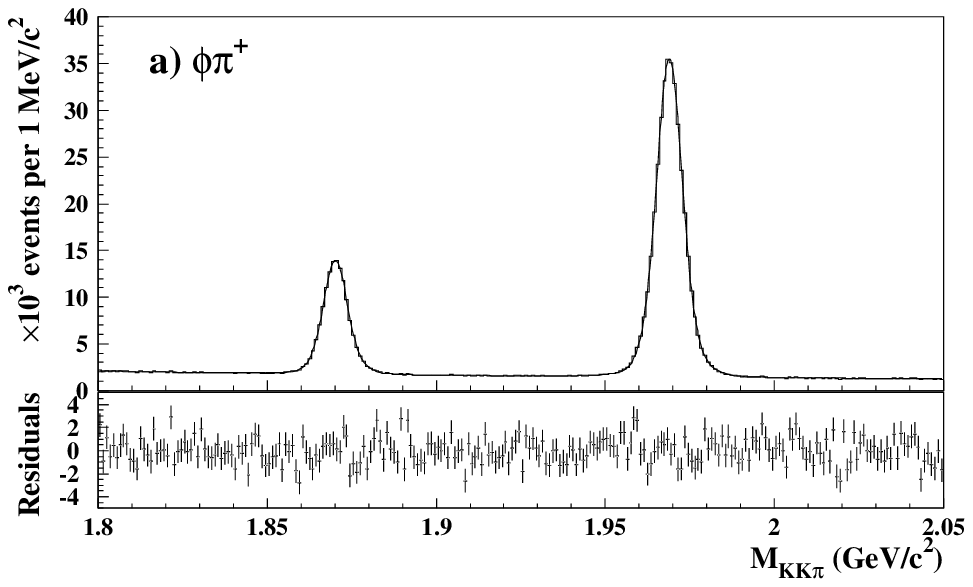}
  \includegraphics[width=8.5cm]{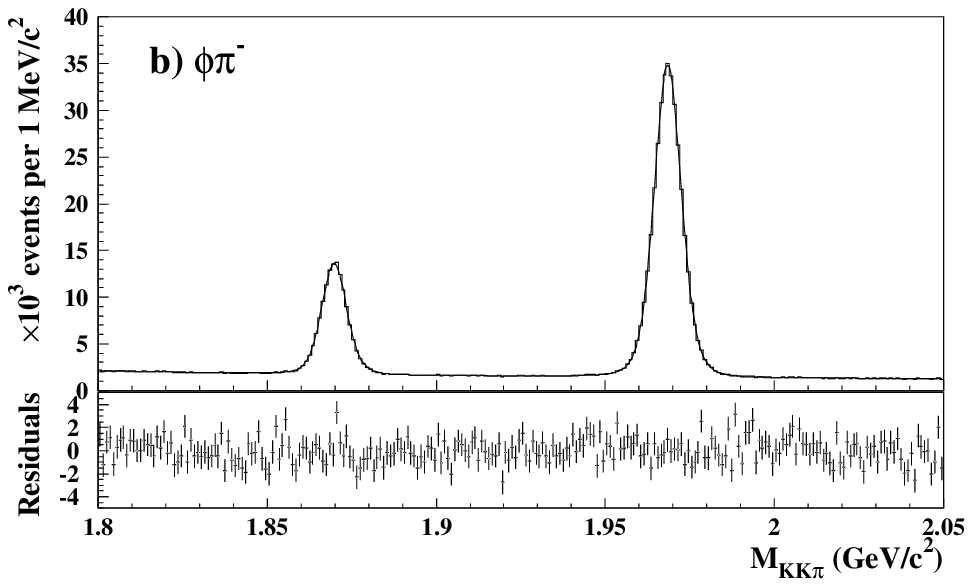}
  \caption{Sum of invariant mass distributions 
    with the sum of fitted functions superimposed for (a) positively and 
    (b) negatively charged $D$ mesons. The plots beneath the distributions 
    show the residuals.}
  \label{fig2}
\end{figure}

Of the approximately 700 sufficiently populated invariant mass distributions 
($D_{(s)}^+$ and $D_{(s)}^-$ per each 3D bin),
658 are fitted successfully. The quality of fits is good: the mean of the normalized $\chi^2$
distribution is 1.000 and the r.m.s is 0.090 for 242 degrees of freedom;
the corresponding confidence level distribution is
uniform. From the fitted yields in 3D bins we calculate the $D^+$ and $D_s^+$ asymmetries, 
and the asymmetry differences $\Delta A_{\rm rec}$.
We consider only those bins in which the yield has a significance greater than $3\sigma$; 
this requirement must be fulfilled for all four measured yields in a
bin. The sum of mass distributions in these bins is shown in 
Fig.~\ref{fig2}. We find $237525\pm 577$ $D^\pm$ and $722871\pm 931$
$D_s^\pm$ decays. The residuals of the sum of all the successfully fitted distributions 
do not show any significant structure. 

The asymmetry difference in each bin is then corrected with $\Delta A_\epsilon^{KK}$ for that bin:
\begin{equation}
  \label{Arec-corr.eq}
  \Delta A_{\rm rec}^{\rm cor} = \Delta A_{\rm rec} - \Delta A_\epsilon^{KK}.
\end{equation}
The corrections are determined using Eq.~(\ref{A_KK.eq}) and the experimental data for the
$P_1(x)$ and $P_2(x)$ distributions. In particular, we calculate $A_\epsilon^{KK}$ 
for events in the signal window, $m_D \pm 15~{\rm MeV}/c^2$, and subtract the asymmetry
for events in an equal width sideband displaced
$\pm 20~{\rm MeV}/c^2$ from the nominal $D$ meson mass $m_D$.

The kaon asymmetry $A_\epsilon^K$,
which is needed in Eq.~(\ref{A_KK.eq}), is measured using $D_s^+ \to \phi \pi^+$ and 
$D^0 \to K^-\pi^+$ decays; for the latter decay the measured asymmetry can be expressed
as $A_{\rm rec} = A_{CP} + A_{FB} - A_\epsilon^K + A_\epsilon^\pi$. By assuming negligible $CP$ 
violation (both are CF decays) and the same forward-backward asymmetry, and by neglecting the 
$A_\epsilon^{KK}$ term in Eq.~(\ref{Arec.eq}), the difference of measured asymmetries is
equal to $A_\epsilon^K$~\cite{iterative}. 

The procedure is similar to that used in~\cite{BRko}.
First we determine the asymmetry of $D_s^+ \to \phi \pi^+$ in 3D bins
using the fitted yields.
This asymmetry map is used to weight $D^0 \to K^-\pi^+$ events in order to determine the
$D^0/\overline{D}^0$ corrected yields in bins of the kaon phase space. The yields
are obtained by the sideband subtraction method as in~\cite{StaricPLB}. 
The kaon asymmetry map is then calculated from
the corrected $D^0/\overline{D}^0$ yields in the range
$0 < p_K < 4~{\rm GeV}/c$ and $-1 < \cos \theta_K < 1$ divided into 10\x 10 equal size
bins.

The following corrections are obtained with Eq.~(\ref{A_KK.eq}) 
for the total 3D phase space:  
$A_\epsilon^{KK}=(+0.060\pm 0.013)\%$ for $D^+$,
$A_\epsilon^{KK}=(-0.051\pm 0.012)\%$ for $D_s^+$ 
and $\Delta A_\epsilon^{KK}=(+0.111\pm 0.025)\%$; the difference is not zero because of 
the opposite signs of the momentum asymmetries in $D^+$ and $D_s^+$ decays, 
as shown in Fig.~\ref{fig1}. 
The uncertainties are due to statistical variations of $A_\epsilon^K$
and $P_1(x) - P_2(x)$; the error on $\Delta A_\epsilon^{KK}$ is included
in the systematic uncertainty.

\begin{figure}[t]
  \includegraphics[width=4.25cm]{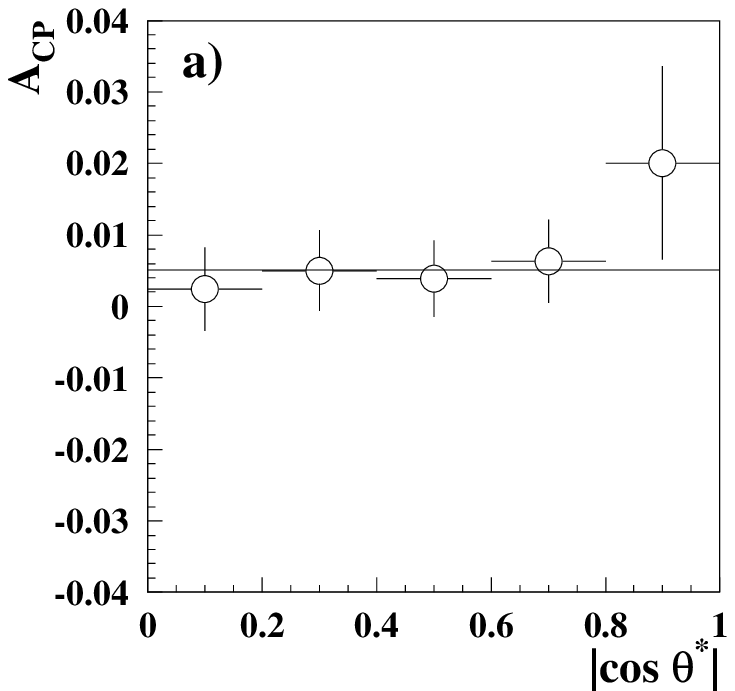}
  \includegraphics[width=4.25cm]{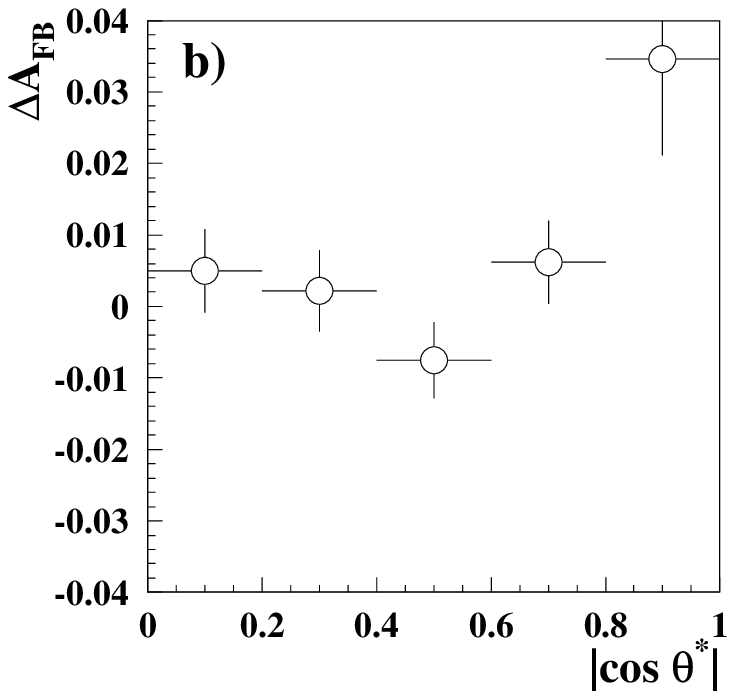}
  \caption{$CP$-violating asymmetry (a) and forward-backward asymmetry
    difference (b) in bins of $|\cos \theta^*|$. The horizontal line in (a) is 
    a constant fit to the data points.}
  \label{fig3}
\end{figure}

The corrected asymmetry differences in 3D bins, $\Delta A_{\rm rec}^{\rm cor}$ defined by 
Eq.~(\ref{Arec-corr.eq}), are used to calculate error-weighted averages 
in bins of $\cos \theta^*$; error-weighted averages are obtained with the least squared fit.
Finally, $A_{CP}^{D^+ \to \phi \pi^+}$ and  $\Delta A_{FB}$ are extracted by adding/subtracting
the asymmetry difference in opposite bins of $\cos \theta^*$:
\begin{eqnarray}
  \label{Acp-extr}
  A_{CP}^{D^+ \to \phi \pi^+}=
  \frac{\Delta A_{\rm rec}^{\rm cor}(\cos\theta^*)+\Delta A_{\rm rec}^{\rm cor}(-\cos\theta^*)}{2},\\
  \label{Afb-extr}
  \Delta A_{FB}=
  \frac{\Delta A_{\rm rec}^{\rm cor}(\cos\theta^*)-\Delta A_{\rm rec}^{\rm cor}(-\cos\theta^*)}{2}.
\end{eqnarray}
The results are shown in Fig.~\ref{fig3}. By fitting the data points of 
Fig.~\ref{fig3}a with a constant we obtain $A_{CP}^{D^+ \to \phi \pi^+}=(0.51 \pm 0.28)\%$, 
where the error is statistical only. The result is consistent with zero within 1.8
standard deviations.  

Figure~\ref{fig3}b shows the difference in forward-backward asymmetries.
The $\chi^2$ test with respect to $\Delta A_{FB}=0$ gives
$\chi^2/{\rm ndf} = 10.57/5$, which corresponds to a confidence level of 6\%;
no significant difference is found between the forward-backward asymmetries for $D^+$ and $D_s^+$. 
A fit to a constant yields a value of $\Delta A_{FB} = (0.25\pm 0.28(stat.))\%$.

We consider five significant sources of systematic uncertainties (Table~\ref{syst_err}).
As discussed before, the $A_\epsilon^{KK}$ corrections are uncertain to 0.025\%. 
The impact of 3D binning 
is studied by changing the binning from  10\x 10\x 10 bins to 20\x 10\x 10, 10\x 20\x 10 and 
10\x 10\x 20 bins; we obtain a 0.026\% variation in $A_{CP}^{D^+ \to \phi \pi^+}$.
By doubling the number of bins in the invariant mass histograms a variation of 0.022\% is
obtained. The impact of signal parameterization is studied by replacing the four-Gaussian shape 
with a triple Gaussian shape and the impact of background parameterization is studied by
replacing the default parametrization 
with a simple exponential function. We also vary the range in which we fit
the distributions; all these changes give a 0.013\% variation in the result. The uncertainty
of the width ratio $f$, which is fixed in the fit, propagates into a 0.012\% uncertainty in
the result. By adding the last two numbers in quadrature we obtain an estimate of 0.018\% 
for the systematic uncertainty of the fitting procedure. 
The last source is the selection of fit results; 
by changing the requirement from $N/\sigma_N>3$ to $N/\sigma_N>5$ we obtain a 0.020\% 
variation in the result. We estimate the total systematic uncertainty
by summing individual contributions in quadrature; we obtain 0.050\%.

\begin{table}
  \caption{\label{syst_err} Summary of systematic uncertainties in $A_{CP}^{D^+ \to \phi \pi^+}$}
  \begin{ruledtabular}
    \begin{tabular}{lc}
      Source & Uncertainty (\%) \\
      \hline
      $A_\epsilon^{KK}$ corrections & 0.025 \\
      3D binning              & 0.026 \\
      Invariant mass binning  & 0.022 \\
      Fitting procedure       & 0.018 \\
      Selection of fit results   & 0.020 \\
      \hline
      Sum in quadrature       & 0.050 \\
    \end{tabular}
  \end{ruledtabular}
\end{table}

In summary, we searched for $CP$ violation in the decays $D^+ \to \phi \pi^+$ 
by measuring the $CP$ violating asymmetry difference between Cabibbo-suppressed ($D^+$) and 
Cabibbo-favored ($D_s^+$) decays in a mass region around the $\phi$ resonance, 
$m_\phi \pm 16~{\rm MeV}/c^2$.
We have made no attempt to disentangle the $\phi$ from other intermediate resonances 
in this mass region. Using 955~fb$^{-1}$ of experimental data 
collected with the Belle detector and assuming negligible CPV in CF decays we measure:
\begin{equation}
  \label{result}
  A_{CP}^{D^+ \to \phi \pi^+} = (+0.51 \pm 0.28 \pm 0.05)\%.
\end{equation}
The result shows no evidence for $CP$ violation and agrees with SM predictions. 
Previously, the most precise measurements were from CLEO~\cite{CLEO} and 
BaBar~\cite{babar}; our measurement improves the precision by more than a factor of five.
We also measure for the first time the difference in the forward-backward asymmetries 
of $D^+$ and $D_s^+$ mesons and find no significant deviation from zero.

\begin{acknowledgments}
  We thank the KEKB group for excellent operation of the
  accelerator, the KEK cryogenics group for efficient solenoid
  operations, and the KEK computer group and
  the NII for valuable computing and SINET3 network support.  
  We acknowledge support from MEXT, JSPS and Nagoya's TLPRC (Japan);
  ARC and DIISR (Australia); NSFC (China); MSMT (Czechia);
  DST (India); MEST, NRF, NSDC of KISTI, and WCU (Korea); MNiSW (Poland); 
  MES and RFAAE (Russia); ARRS (Slovenia); SNSF (Switzerland); 
  NSC and MOE (Taiwan); and DOE (USA).
\end{acknowledgments}


\begin{thebibliography}{99}
\bibitem{ch_conj} Unless noted otherwise, charge-conjugate processes are
  implicitly included throughout this Letter.
\bibitem{CPV_dir} In contrast to charged $D$ mesons, neutral mesons can
  exhibit CPV in decays as well as a consequence of mixing; while the
  former $CP$ asymmetries can be different in various decay modes, the
  mixing-induced CPV is decay mode independent.
\bibitem{iterative} Corrections for a non-zero $A_\epsilon^{KK}$ can be applied iteratively,
  using $A_\epsilon^K$ from the previous iteration and Eq.~(\ref{A_KK.eq}). 
  However, the asymmetry map is found to change negligibly after the iterations converge.
\end{thebibliography}
\end{document}